\documentclass{aa}  

\usepackage{natbib,twoopt}
\usepackage{multirow}
\usepackage{amsmath}% Advanced maths commands
\usepackage{amssymb} % Extra maths symbols
\usepackage[T1]{fontenc}
\usepackage{blindtext}
\usepackage{hyperref}
\hypersetup{
    colorlinks=true,
    linkcolor=red,
    filecolor=blue,
    citecolor=blue,
    urlcolor=blue,
    pdftitle={Overleaf Example},
    pdfpagemode=FullScreen,
}
\usepackage{graphicx}
\usepackage{gensymb}
\usepackage{subcaption}
\usepackage{pythonhighlight}
\usepackage{listings}
\usepackage{xcolor}
\usepackage{multirow}
\usepackage{siunitx}
\usepackage{orcidlink}

\usepackage{amsmath} % or simply amstext

\usepackage[varg]{txfonts}

\usepackage{graphicx}	% Including figure files
\usepackage{amsmath}	% Advanced maths commands

\newcommand{\relxill}{\textsc{relxill}}
\newcommand{\xillver}{\textsc{xillver}}
\newcommand{\relxillD}{\textsc{relxillD}}
\newcommand{\xillverD}{\textsc{xillverD}}
\newcommand{\relxillCp}{\textsc{relxillCp}}
\newcommand{\xillverCp}{\textsc{xillverCp}}
\newcommand{\nthcomp}{\textsc{nthcomp}}

\newcommand{\xmm}{\textit{XMM-Newton}}
\newcommand{\nustar}{\textit{NuSTAR}} 
\definecolor{codegreen}{rgb}{0,0.6,0}
\definecolor{codegray}{rgb}{0.5,0.5,0.5}
\definecolor{codepurple}{rgb}{0.58,0,0.82}
\definecolor{backcolour}{rgb}{0.95,0.95,0.92}

\lstdefinestyle{mystyle}{
  backgroundcolor=\color{backcolour}, commentstyle=\color{codegreen},
  keywordstyle=\color{magenta},
  stringstyle=\color{codepurple},
  basicstyle=\ttfamily\footnotesize,
  breakatwhitespace=false,         
  breaklines=true,                 
  captionpos=b,                    
  keepspaces=true,                                               
  showspaces=false,                
  showstringspaces=false,
  showtabs=false,                  
  tabsize=2
}

\begin{document}

\title{The origin of the soft excess in the luminous quasar HE 1029-1401}

\author{B. Vaia \orcidlink{0000-0003-0852-0257}\inst{1,2,3}
        , F. Ursini\inst{4}, G. Matt\inst{4}, D.R. Ballantyne\inst{5}, S. Bianchi\inst{4}, A. De Rosa\inst{6}, R. Middei\inst{7,8}, P.O. Petrucci\inst{9},\\ E. Piconcelli\inst{7} \and A. Tortosa\inst{7}
          }
          
\authorrunning{B. Vaia et al.}

\institute{Scuola Universitaria Superiore IUSS Pavia, Piazza della Vittoria 15, 27100 Pavia, Italy\\ \email{beatrice.vaia@iusspavaia.it}
  \and Department of Physics, University of Trento, Via Sommarive 14, 38123 Povo (TN), Italy 
  \and Istituto Nazionale di Astrofisica, Istituto di Astrofisica Spaziale e Fisica Cosmica di Milano, via A. Corti 12, 20133 Milano, Italy
   \and Dipartimento di Matematica e Fisica, Università degli Studi Roma Tre, via della Vasca Navale 84, 00146, Roma, Italy
   \and Center for Relativistic Astrophysics, School of Physics, Georgia Institute of Technology, 837 State Street, Atlanta, GA 30332-0430, USA
   \and INAF Istituto di Astrofisica e Planetologia Spaziali, Via del Fosso del Cavaliere 100, I-00133 Roma, Italy
   \and INAF - Osservatorio Astronomico di Roma, Via Frascati 33, I-00040 Monte Porzio Catone, Italy
   \and Space Science Data Center, SSDC, ASI, Via del Politecnico snc, I-00133 Roma, Italy
   \and Univ. Grenoble Alpes, CNRS, IPAG, 38000 Grenoble, France}
   \date{Received; accepted}

% \abstract{}{}{}{}{} 
% 5 {} token are mandatory

\abstract{
The enigmatic and intriguing phenomenon of the "soft excess" observed in the X-ray spectra of luminous quasars continues to be a subject of considerable interest and debate in the field of high-energy astrophysics. This study focuses on the quasar HE 1029-1401 ($z=0.086$, $\log(L_{\rm{bol}}/[\rm{erg\,s^{-1}}])= 46.0 \pm 0.2$), with a particular emphasis on investigating the properties of the hot corona and the physical origin of the soft excess.
In this study, we present the results of a joint \textit{XMM-Newton}/\textit{NuSTAR} monitoring campaign of this quasar conducted in May 2022. The source exhibits a cold and narrow Fe $\rm{K}\alpha$ emission line at 6.4 keV, in addition to the detection of a broad component.
Our findings suggest that the soft excess observed in HE 1029-1401 can be adequately explained by Comptonized emission originating from a warm corona. Specifically, fitting the spectra with two \nthcomp\, component we found that the warm corona is characterized by a photon index ($\Gamma^{w}$) of  $2.75\pm0.05$ and by an electron temperature ($kT_{e}^{w}$) of $0.39^{+0.06}_{-0.04}$ keV, while the optical depth ($\tau^{w}$) is found to be $23\pm3$.
We also test more physical models for the warm corona, corresponding to two scenarios: pure Comptonization and Comptonization plus reflection. Both models provide a good fit to the data, and are in agreement with a radially extended warm corona having a size of a few tens of gravitational radii.}


%\titlerunning
\maketitle
%\authorrunning
%%%%%%%%%%%%%%%%% BODY OF PAPER %%%%%%%%%%%%%%%%%%
\nolinenumbers
\section{Introduction}

Active Galactic Nuclei (AGN) are highly energetic astrophysical sources powered by the accretion of matter onto a supermassive black hole (SMBH). AGN emit radiation across the whole electromagnetic spectrum. The accretion disk is the primary source of optical/UV emission, while X-ray emission is the result of the inverse Compton scattering of disk optical/UV photons by the hot thermal distribution of electrons that constitute an optically thin corona.
This physical mechanism explains the observed power-law shape and the high-energy cut-off often observed at 100 – 150 keV \citep{Malizia2014,Fabian,Tortosa}. The reflection of the primary coronal emission by surrounding material can produce a reflection continuum characterized by a Compton hump peaked at 20 – 30 keV \citep[e.g.][]{George1991,Ricci18} and fluorescence emission lines from heavy elements, in particular neutral/ionized Fe K emission is ubiquitous in AGN spectra.

The soft X-ray excess is frequently observed in the spectra of unobscured AGN (e.g. \citealt{1993Walter,2004Page,Bianchi2009,2020Gliozzi}). 
Currently, the most debated scenarios for the origin of the soft excess are the two-corona model and the model involving relativistic ionized reflection \citep[e.g.][]{Crummy2006,Walton2013,2016Garcia}. The first involves a warm optically thick corona and a hot optically thin corona, with the soft excess originating from the Comptonization of optical/UV photons from a warm electron corona with values of optical depth ($\tau^{w}$) and electron temperature ($kT^{w}$) being, respectively, $10-20$ and $\simeq 1 \rm{keV}$ (e.g. \citealt{1998Mag,2011Med,2012Jin,2015Med,Petrucci1,2018Porquet}, and references therein). The second model explains the soft excess as the reprocessing of the primary continuum off an ionised disk \citep{Ross}, with the soft excess profile being due to the overlapping of blurred soft X-ray emission lines.

In this paper, we investigate the properties of the hot corona and the physical origin of the soft excess in the radio-quiet quasar HE 1029-1401. 
HE 1029-1401 ($z=0.086$ \citealt{Wis}) is a bright quasar hosting a SMBH of mass  $\log(M_{\rm{BH}}/M_{\odot})=8.7 \pm 0.3$ \citep{Husemann}.
HE 1029-1401 was discovered in the 1980s by the Edinburgh-Cape Blue Object (EC) Survey with the UK-Schmidt telescope at the Anglo Australian Observatory (AAO) but was confirmed as a quasar only thanks to the Hamburg/ESO Survey which investigated the southern extragalactic sky in 1990 \citep{Wis}. With an apparent magnitude V = 13.6, at the time of discovery it was the brightest quasar ever found in the optical. The source has a bolometric luminosity $\log(L_{\rm{bol}}/[\rm{erg\,s^{-1}}])= 46.0 \pm 0.2$ corresponding to an Eddington rate of $\log({L_{\rm{bol}}/L_{\rm{Edd}}}) = -0.9 \pm 0.2$ \citep{Husemann}.
The first detection of the X-ray emission of this AGN occurred thanks to the Japanese satellite Ginga. The X-ray spectrum studied by \citet{Iwasawa}, showed a flat power law and an excess of emission in the soft band. More recently, HE 1029-1401 has been included in the sample studied by  \citet{Petrucci1}; the authors used a large sample of radio-quiet AGN, observed by \textit{XMM-Newton}, to test the possible origin of the soft excess from a warm corona. The model provides a good fit to the \textit{XMM-Newton} data, however the lack of high-energy coverage does not allow for tight constraints on the reflection component. This paper reports on the analysis of a simultaneous \textit{XMM-Newton} and \textit{NuSTAR} observation performed in May 2022.

The paper is structured as follows: Sect. 2 describes the observations and data reduction, Sect. 3 presents the analysis of the XMM–Newton and \textit{NuSTAR} spectra, first focused on the hard X-ray band and then extended to the broad X-ray band. In Sect. 4 we discuss the results and we present our conclusions. 

\section{Observations and data reduction}

%More recently, HE 1029-1401 has been included in the group of sources studied by  \citet{Petrucci1}; the authors used a large group of AGNs (observed with \textit{XMM-Newton}) to test the possible origin of the soft excess from a warm corona.\\
%In particular, the model used by the authors includes two \emph{nthcomp} \citep{Zdz, Zycki} components (to represent the comptonization continuum generated by the two coronae) a single reflection component (the non-relativistic component) and a component to represent the contribution of the BLR (the study in fact, it also included the optical monitor data).\\ 
%The model provides a good fit to the \textit{XMM-Newton} data, however the lack of high-energy coverage does not allow for tight constraints on the reflection component.\\

\begin{table}
	\centering
	\caption{Obs. ID, start date and exposure time for HE 1029-1401 \textit{XMM-Newton} and \textit{NuSTAR} observations.}
	\label{tab:1}
	\begin{tabular}{lccl} % four columns, alignment for each
		\hline
		Satellite & Obs. ID & Start time (UTC) & Net exp.(s)\\
		\hline
        \textit{XMM-Newton} &  0890410101 & 2022-05-23 & $7.14$ x $10^4$\\
        \textit{NuSTAR} & 60701046002 & 2022-05-23 & $2.24$ x $10^5$\\
    	\hline
	\end{tabular}
\end{table}

\textit{XMM-Newton} observed the source with the optical monitor (OM; \citealt{Mason}) and the EPIC cameras \citep{Struder,Turner}. 
We processed the data using the \textit{XMM-Newton} Science Analysis System (SAS v20). The OM photometric filters U, UVW1, UVM2, and UVW2 were used, in the image mode, for a total exposure time of 21 ks each. We processed the OM data with the SAS pipeline omichain\footnote{
The standard omichain task applies flat-field corrections, detects sources, computes source positions and their count rates, and applies the proper calibration to convert to instrumental magnitudes. HE~1029-1401 is by far the brightest source in the field of view, and its count rates are extracted from a region of 5.3 arcsec.
}, and converted to OGIP format with the SAS task om2pha. 
The EPIC-pn and MOS cameras operated in the Small Window mode meaning a live time of 71\%. We use only pn data for the spectral analysis, because of the higher effective
area compared with MOS. Moreover, MOS data suffer from low-energy noise, leading to significant residuals below 1-2 keV. However, we verified that the spectral parameters in the 2--10 keV band are consistent among MOS and pn, even though with larger uncertainties in MOS. No significant pile-up is foud using the SAS task epatplot. We determined source extraction radii and screening for high-background intervals via an iterative process that maximizes the signal-to-noise ratio, following \citet{Piconcelli}. We extracted the background from circular regions with a radius of 50 arcsec. Source radii were allowed to be in the range 20–40 arcsec, and the best radius was found to be 40 arcsec. 
To generate the effective area, we used the SAS task arfgen with the keyword applyabsfluxcorr=yes, which improves the agreement with the \nustar\ spectra. 
Finally, the pn spectra were grouped to have at least 30 counts in each bin, and not oversampling the spectral resolution by a factor greater than 3.

We reduced the \textit{NuSTAR} data with the standard pipeline (nupipeline) of the \textit{NuSTAR} Data Analysis Software (nustardas, v1.9.7), using calibration files from \textit{NuSTAR} caldb v20221229. We extracted spectra and light curves with the standard tool nuproducts for the two detectors in focal plane modules A and B (FPMA and FPMB). We extracted the background data from circular regions with a radius of 75 arcsec, while the source radii were calculated to maximize the signa-to-noise like for \textit{XMM-Newton}. The final source radii were 80 arcsec. We regrouped the spectra with the tool ftgrouppha, part of HEASOFT v6.31.1, according to the optimal scheme of \citet{KaastraeB} with the additional requirement of a minimum signal-to-noise of 3 in each bin. The spectra from FPMA and FPMB were analysed jointly but not co-added. 
To account for the cross-calibration between the two \textit{NuSTAR} modules, we included a constant term in the spectral fits. We find a difference of 2\% between the two modules. The log of the data sets analysed in this paper is listed in Table~\ref{tab:1}.

\begin{figure}
	\includegraphics[width=\columnwidth]{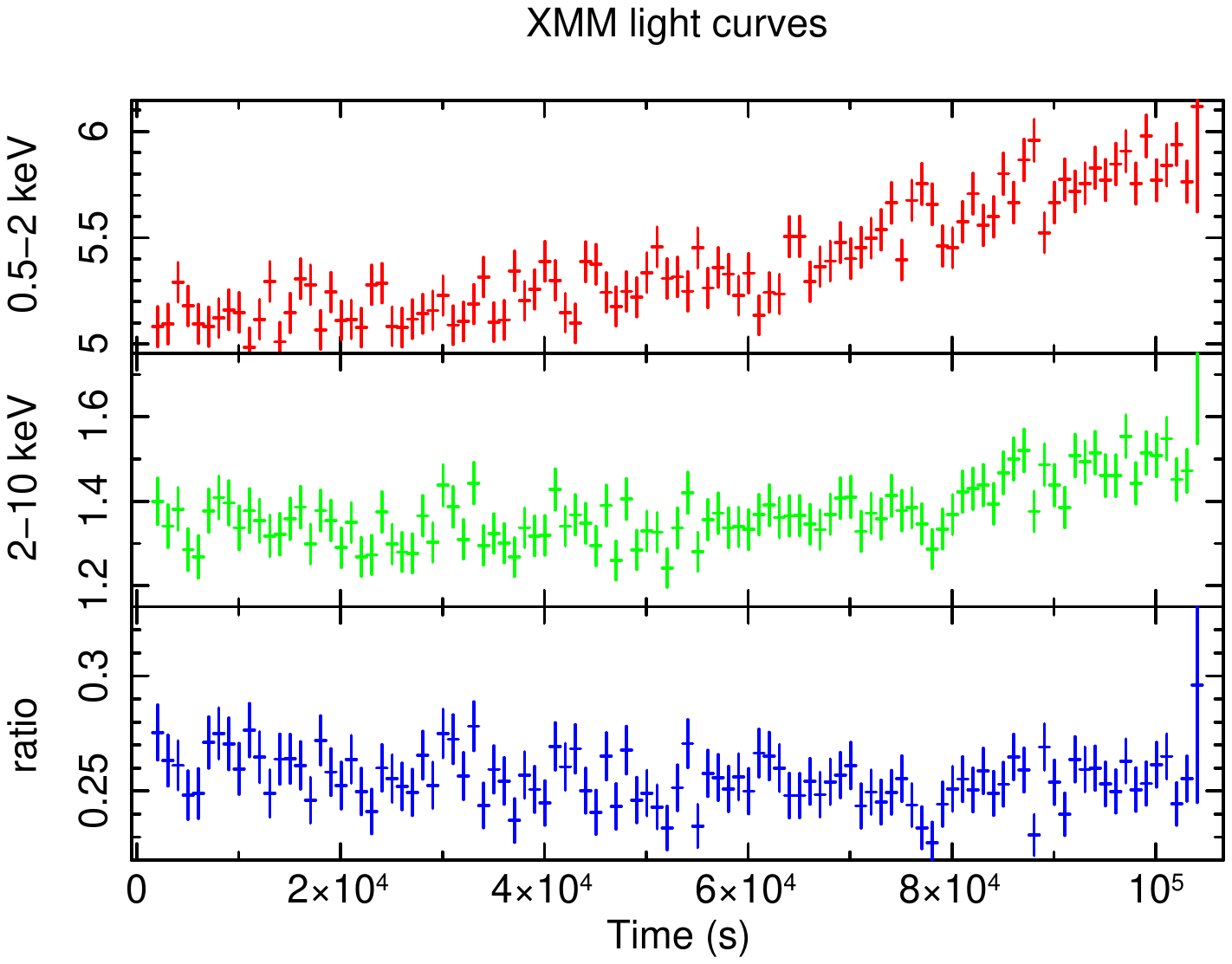}
    \caption{\textit{XMM} count rate light curves. Top panel: light curve for the energy interval $0.5-2$ keV . Middle panel: light curve for the energy interval $2-10$ keV . Bottom panel: ratio between the two energy band.}
    \label{fig:example_figureXMM}
\end{figure}

\begin{figure}
	\includegraphics[width=\columnwidth]{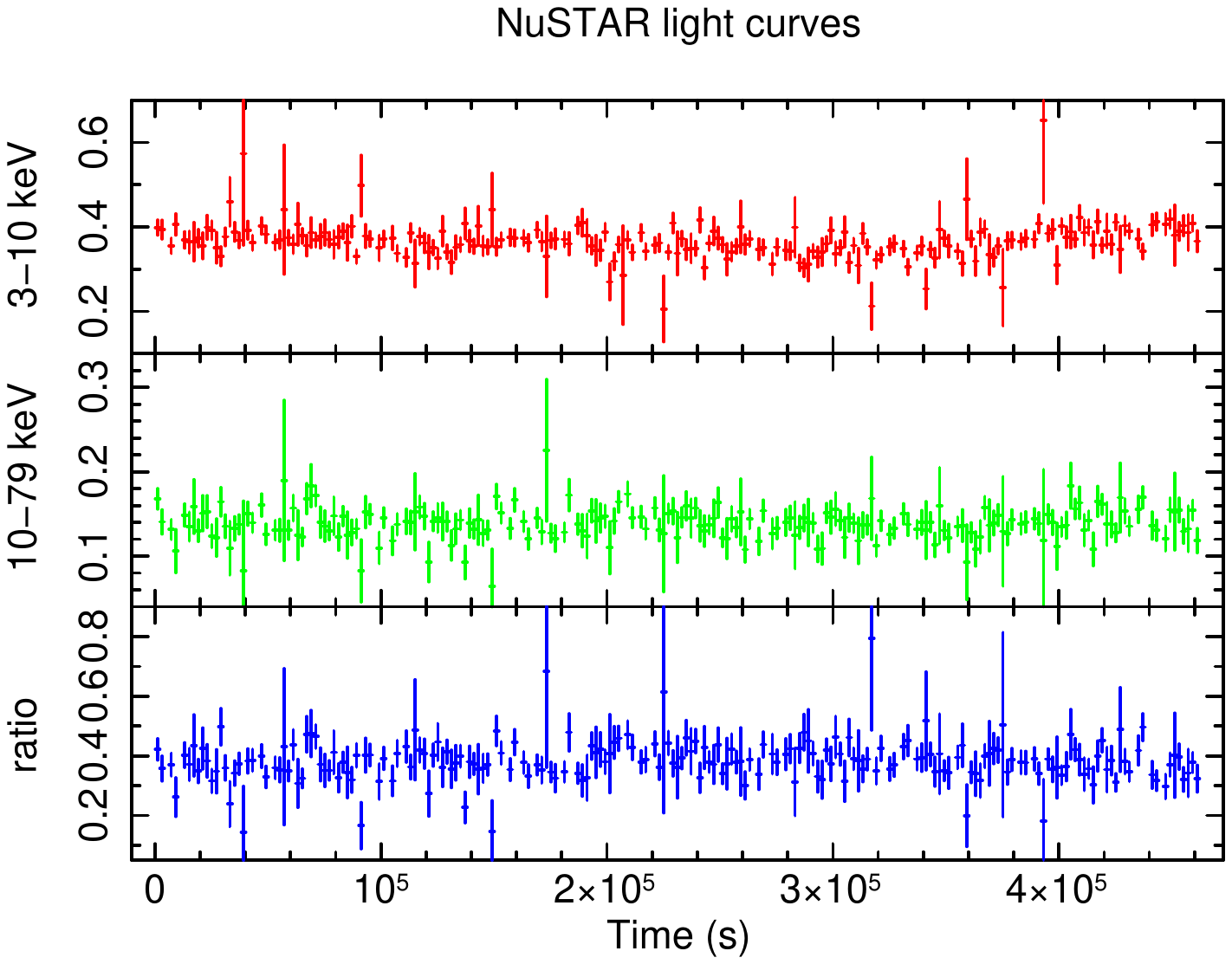}
    \caption{\textit{NuSTAR}/FPMA background-subtracted count rate light curves. Top panel: light curve for the energy interval $3-10$ keV . Middle panel: light curve for the energy interval $10-79$ keV. Bottom panel: ratio between the two energy band.}
    \label{fig:example_figureNus}
\end{figure}

%\section{Timing properties}
The \textit{XMM-Newton} and \textit{NuSTAR} light curves of HE 1029-1401 are plotted in Figure~\ref{fig:example_figureXMM} and Figure~\ref{fig:example_figureNus} respectively. The light curves shows an X-ray variability, but no hardness ratio variation is observed.
The observation occurred in a low flux period ($\simeq 1 \times 10^{11}\,\rm{erg}\,\rm{s}^{-1}\,\rm{cm}^{-2}$) and for \textit{NuSTAR} the background was particularly high dominating over the source at energies greater than 30 keV. For these reasons, we exclude the \textit{NuSTAR} data above 30 keV for the spectral analysis.

\section{Spectral analysis}
We performed the spectral analysis with the \emph{xspec} v12.13.0 package \citep{Arnaud}.
Fits were performed on the rebinned pn and \textit{NuSTAR} spectra plus the OM photometric data, using the $\chi^2$ minimization technique.
All errors are quoted at the 90\% confidence level ($\Delta\chi^2=2.71$) for one interesting parameter. In all of our fits, we included neutral absorption (\textsc{tbabs} model in \emph{xspec}) from Galactic hydrogen with column density $N_{\rm{H}}=5.74\rm{x}10^{20}\,\rm{cm}^{-2}$ \citep{Kalberla}.
 The cosmological parameters $\rm{H_0}=70\,\rm{km\,s^{-1}\,Mpc^{-1}}$, $\Omega_{\Lambda}=0.73$ and $\Omega_m= 0.27$ are adopted.

\subsection{The reflection component}

\begin{figure}
	\includegraphics[width=\columnwidth]{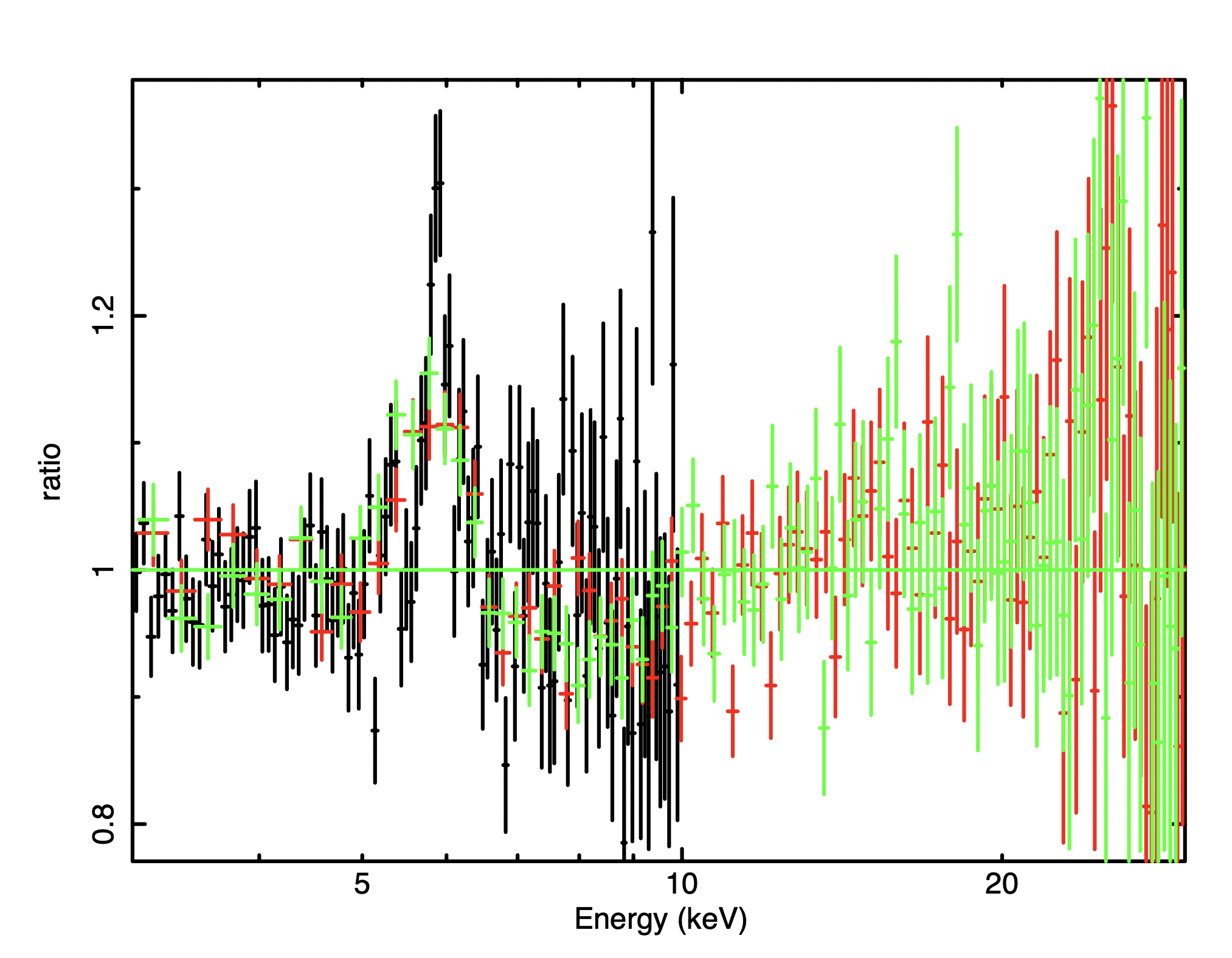}
    \caption{Data/model ratio for the \textit{XMM-Newton} and \textit{NuSTAR} spectrum in the 3-30 keV band using the model: \textsc{const × tbabs × cutoffpl}.}
    \label{ratio}
\end{figure}

We started by investigating the Fe K$\alpha$ line at $6.4$ keV.
Initially, we constructed the spectra solely based on a power law model. Figure \ref{ratio} illustrates the ratio between the observed data and the model derived from this initial fit. It reveals an amplification in the data/model ratio up to 5 keV, along with an emission line at 6.4 keV. To address this, we introduced a single Gaussian line with a fixed energy of 6.4 keV and a fixed sigma of zero; in this first part, we analysed only the \xmm/pn spectrum in the 3-10 keV band, given the much better energy resolution compared with \nustar. The resulting $\chi^2/\rm{d.o.f.}$ was 128/104. Subsequently,leaving the intrinsic width free to vary we found a reduced $\chi^2/\rm{d.o.f.}$ of 121/103. Despite the modest enhancement observed, we proceeded with further test. We modeled the data using two emission lines, both located at 6.4 keV, one narrow ($\sigma = 0$) and one broadened due to relativistic effects ($\sigma$ free to vary).
In this case, we found a significant improvement, with $\chi^2/\rm{d.o.f.}=114/102$ ($\Delta \chi^2/\Delta \rm{d.o.f.}= -24/-2$). The broad emission line has $\sigma = 0.38_{-0.16}^{+0.26}$ keV.

%and a probability of chance improvement (calculated with the F-test) of 0.003. 
We concluded that the data are consistent with the presence of both a narrow and a broad emission line. 

%\begin{table}
%\centering
%\label{tab:2}
%\caption{Properties of the Fe $K_{\alpha}$ emission line.} 
%\begin{tabular}{ccc}
%\hline
%$N_{\rm{narrow}} (10^{-6})$ & $\rm{ph}$ $\rm{keV}^{-1}$ $\rm{cm}^{-2}$ $\rm{s}^{-1}$ & $7.13_{-5.09}^{+5.45}$\\
%\hline
%$\sigma$ & keV & $0.38_{-0.16}^{+0.26}$\\
%\rule[-4mm]{0mm}{1cm}
%$N_{\rm{broad}} (10^{-5})$ & $\rm{ph}$ $\rm{keV}^{-1}$ $\rm{cm}^{-2}$ $\rm{s}^{-1}$ & $1.15_{-0.30}^{+0.26}$\\
%\hline
%\end{tabular}
%\label{risultati}
%\end{table}

\begin{figure}
	\includegraphics[width=\columnwidth]{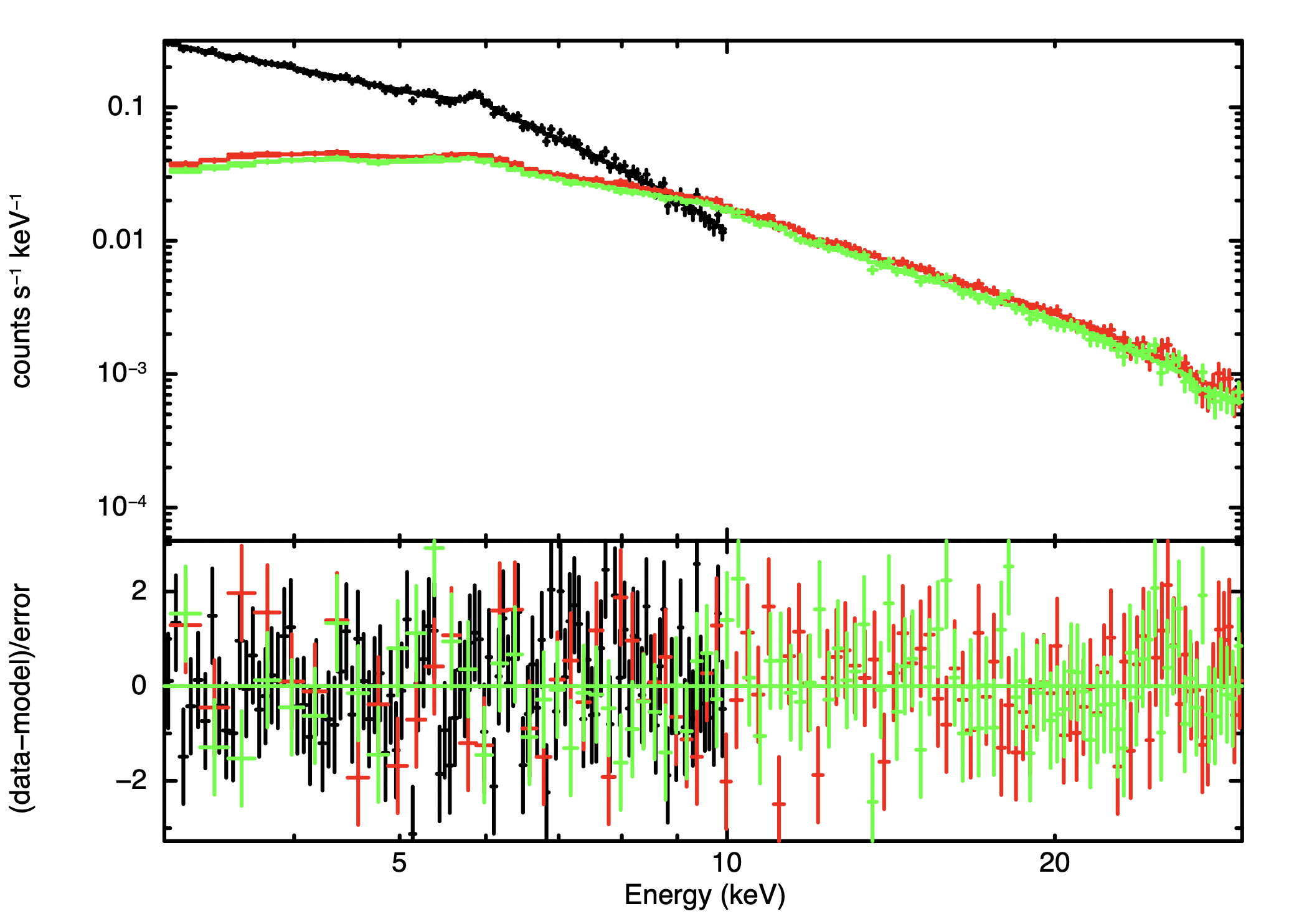}
    \caption{Upper panel: Spectroscopic data and best fit model between 3 and 30 keV. Lower panel: residuals of the model. Model used during the fit: $\textsc{const*tbabs*(\,cutoffpl\,+\,relxill\,+\,xillver)}$. \textit{XMM-Newton} data are in black, the \textit{NuSTAR FMPA} data in red and the \textit{NuSTAR FMPB} data in green.}
    \label{res1}
\end{figure}

%\subsection{Reflection component}
As a next step, we fitted jointly the \xmm/pn and \nustar\ data in
the 3–30 keV band with a self-consistent reflection model. Since we found evidence of both a narrow and a broad component of the Fe K$\alpha$ emission line, we included two different reflection components. Our model consists of a power law with an exponential cut-off ($\rm{E_{cutoff}}$), a non-relativistic reflection component (\xillver\ \citealt{Garcia&Kall,Garcia&Dau}) and a relativistic one (\relxill\ \citealt{Garcia&Dau2014,Dauser}).\footnote{\textsc{tbabs × (cutoffpl+ relxill + xillver)}}
We fixed the source inclination angle at $30^{\circ}$, the outer radius of the accretion disk  to $400\,\rm{R}_{\rm{g}}$ and the spin of the black hole to $0.998$. This leaves us with the freedom to vary the internal radius of the accretion disk, as the internal disk radius and spin parameter are degenerate.
The value of the photon index in \relxill\ - as in \xillver\ - was linked to that of the continuum. The ionization parameter in \xillver\ was fixed at $\log \xi = 0$, while in \relxill\ is left free to vary.
All the other components were free to vary.  With this model, we obtained an acceptable fit with $\chi^2/\rm{d.o.f.}\,=\, 338 / 303$ and the following best-fit parameters: $R_{\rm{int}}$ = $30_{-7}^{+8}\,\rm{R}_{\rm{g}}$, $\rm{A}_{\rm{Fe}}=1.7_{-0.3}^{+0.4}$, $\Gamma^{h}=1.79\pm0.02$, $\rm{E_{cutoff}} = 84_{-12}^{+20}$ keV and $\log(\xi /\rm erg\,\rm cm\,\rm s^{-1})=1.0^{+0.6}_{-0.5}$. In Fig.~\ref{res1} we plot the spectra and the residuals of the model.
%We can conclude that a relativistic reflection is required for the fit and that the source have a not much ionized accretion disk. 
%log xi was free to vary during the first fits and then tied to the best-fit parameter for simplicity.

\subsection{Soft excess and relativistic reflection}

After the analysis of the reflection component, we fitted the whole \textit{XMM-Newton} and \textit{NuSTAR} data in the 0.35-30 keV band. In our case, thanks to the high \textit{NuSTAR} sensitivity in the hard X-rays, we can better constrain the reflection component and the primary continuum at high energies.

\begin{figure}
	\includegraphics[width=\columnwidth]{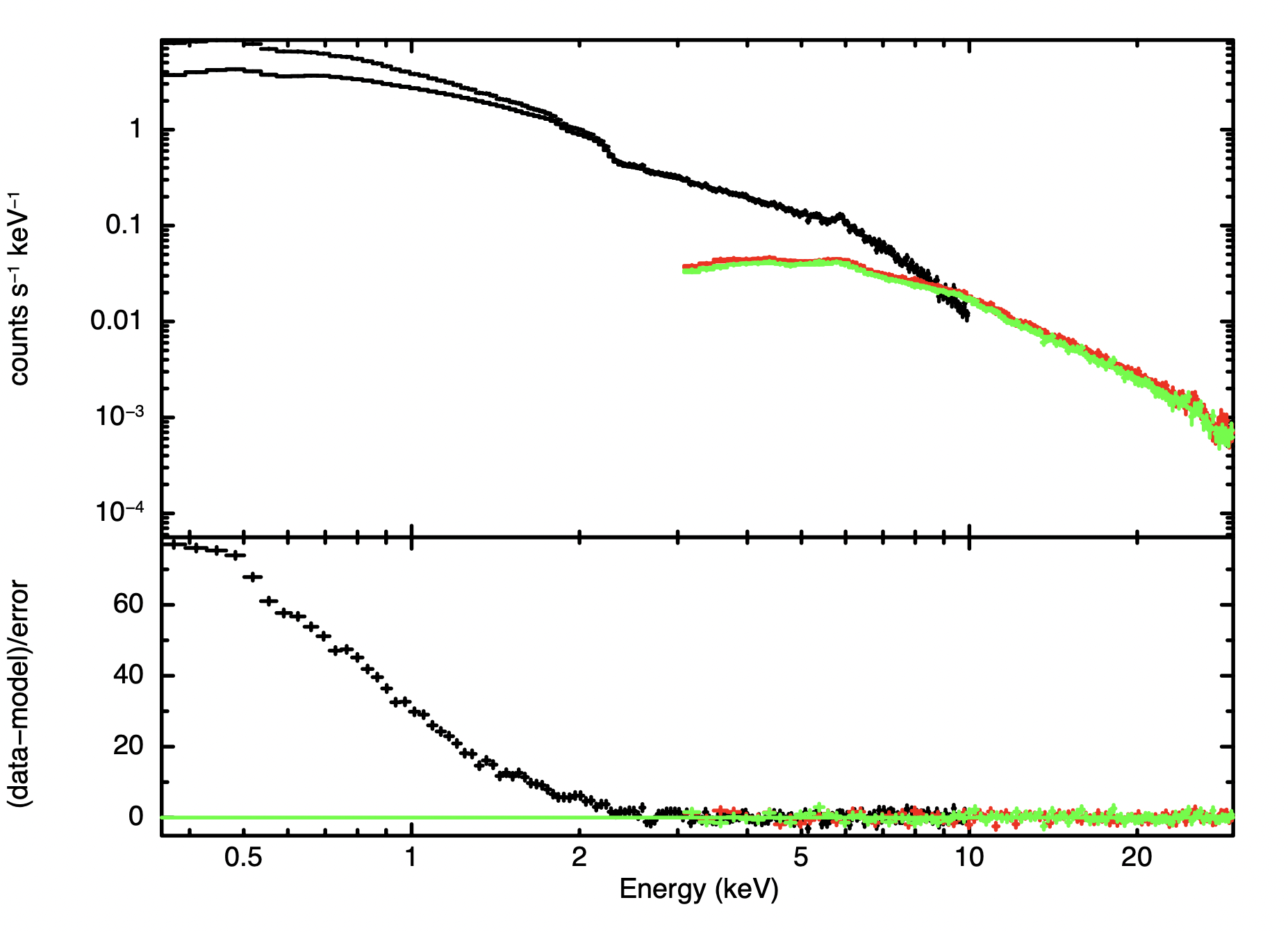}
    \caption{Upper panel: Spectroscopic data between 0.35 and 30 keV and best fit model. Lower panel: residuals of the model. Model used during the fit: \textsc{tbabs × (cutoffpl+ relxill + xillver)}. \textit{XMM-Newton} data are in black, the \textit{NuSTAR FMPA} data in red and the \textit{NuSTAR FMPB} data in green.}
    \label{soft}
\end{figure}

The soft excess is clearly visible in Fig.~\ref{soft}, which represents the data in the entire available energy range and the best fit model discussed in the previous section. We first tested a phenomenological model for the soft excess, namely a disk blackbody component (\textsc{diskbb} in XSPEC). We left free to vary the inner temperature of the disk and the normalization of \textsc{diskbb}. For the temperature, a best fit value of $172\pm7$ eV was found.

Then, we tested whether a model including only reflection components is able to explain the soft excess.

We tested two different flavors of \xillver\ and \relxill. 
In the first case, we used the model $\textsc{const×tbabs×(\,cutoffpl\,+\,relxill\,+\,xillver)}$ (model A). 
We employed \relxill\ to characterize both the continuum and the ionized reflection originating from the inner disc, which generate a soft excess in this model. Additionally, \xillver\ was utilized for the constant reflection component.
In \relxill, the free parameters were the ionization 
%(best fitting value $\log \xi_{i} = 1.56$) 
and the normalization, while the iron abundance is tied to the value of \xillver. The outer disk radius was fixed at $400\,\rm{R}_{\rm{g}}$, while the inner disk radius was left free to vary during the fit. 
%(best fitting value $\rm{R}_{\rm{in}}=10\,\rm{R}_{\rm{g}}$).
The value of the photon index and of the cut-off energy in \relxill\ - as in \xillver\ - were linked to those of the continuum. With this model, we found an unacceptable fit with $\chi^2/\rm{d.o.f.} = 790/369$ and strong residuals below 1 keV. We also tried to include a warm absorber (modeled with a \textsc{cloudy} table, \citealt{2013Ferland} computed assuming a standard type 1 AGN illuminating spectrum) to model the residuals, but no significant improvement was found (even in this case we found a reduced $\chi^2>2$).

%For the first one: $\emph{const*tbabs*(\,cutoffpl\,+\,relxill\,+\,xillver)}$, we found an unacceptable fit with $\chi^2/\rm{d.o.f.} = 818/370$
%and a very soft power law photon index very soft: $2.19$.\\
%Including only the reflection components we found 
%and strong residuals below 1 keV (see Fig.~\ref{res_riflessione}).\\
%In relxill, the parameters free to vary are the ionization (best fitting value $\log \xi_{i} = 1.56$) and the normalization, while iron abundance is tied to the value of xillver model.\\  The outer disk radius was fixed at $400\,\rm{R}_{\rm{g}}$, while the inner disk radius is left free to vary during the fit (best fitting value $\rm{R}_{\rm{in}}=10\,\rm{R}_{\rm{g}}$).The value of the the photon index and the cut-off energy in rellxill - as in xillver - were linked to those of the continuum (as in the previous section, the cut-off energy was fixed at 100 keV).\\ 

\begin{figure}[htp]
\centering
\subfloat{\includegraphics[width = 0.4 \textwidth]{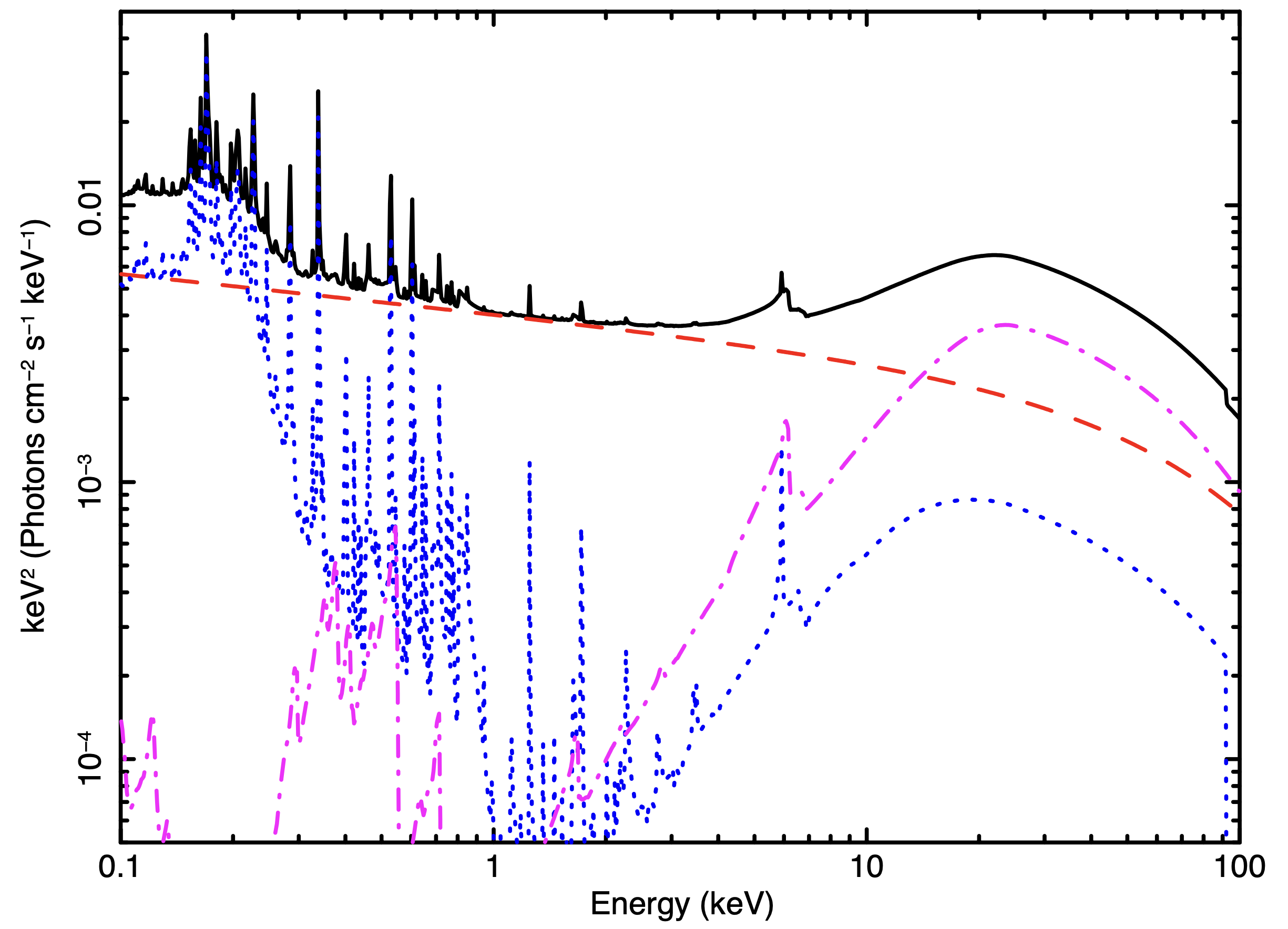}} \label{res_riflessioneD}
\subfloat{\includegraphics[width = 0.4 \textwidth]{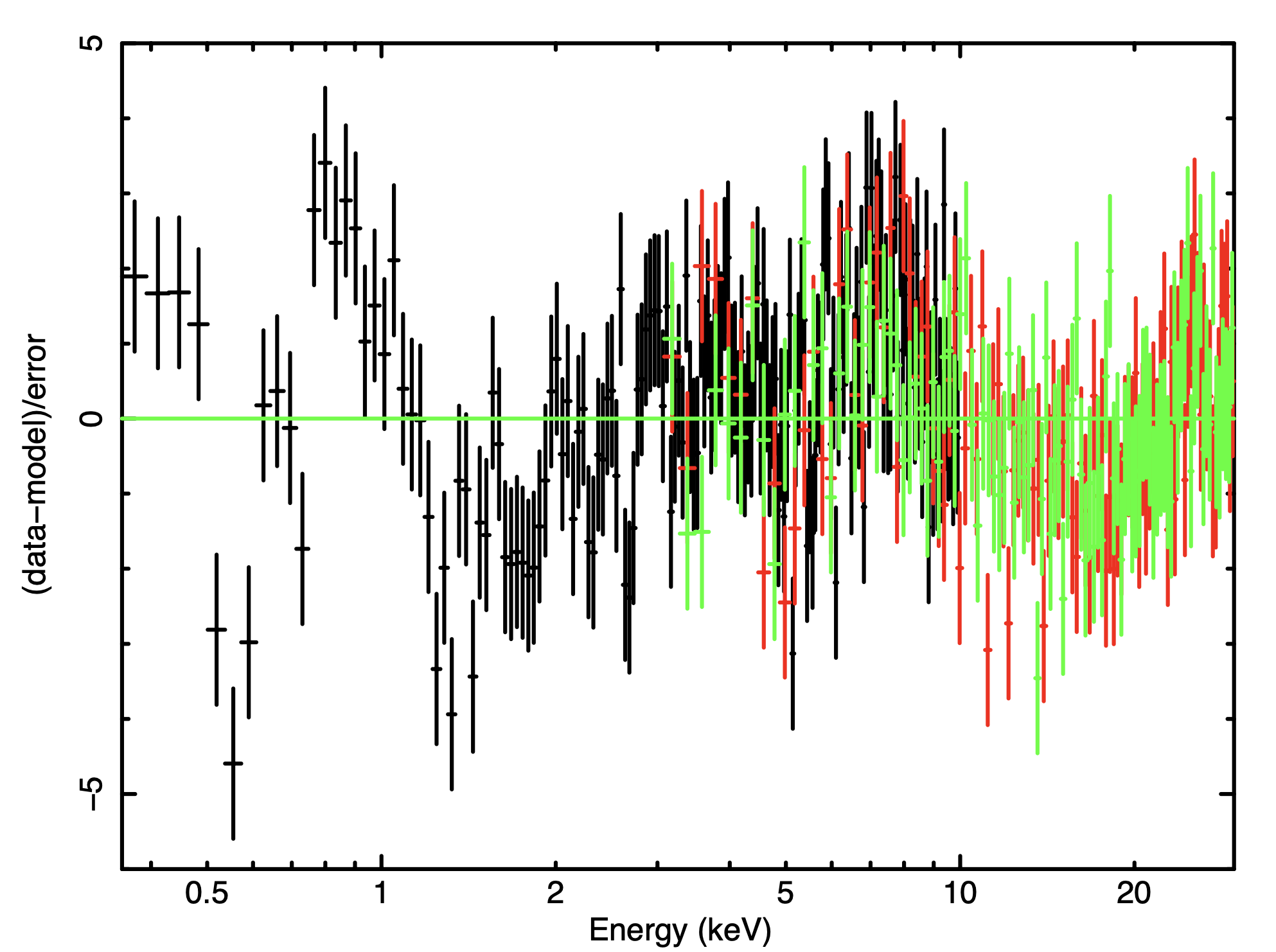}} \label{res_riflessioneD1}
\caption{Upper panel: Model B (\textsc{cutoffpl+relxillD+xillverD}) of the X-ray spectrum of HE 1029-1401. The solid thicker black line is the best-fit model while the dashed lines are the different spectral components: in red the emission of the hot corona, the non-relativistic reflection component in blue and the relativistic reflection in pink.\\ Lower panel: residuals of the model. \textit{XMM-Newton} data are in black, the \textit{NuSTAR FMPA} data in red and the \textit{NuSTAR FMPB} data in green.
Model used during the fit: model B.}
   \label{ress_riflessioneD}
\end{figure}

For the second model (model B) we used \relxillD\ and \xillverD, that includes the density of the reflecting material as a free parameter \citep{2016Garcia}.\footnote{model B: \textsc{const×tbabs×(\,cutoffpl\,+\,relxillD\,+\,xillverD)}}
Free-free absorption is directly proportional to density squared, so an uptick in density results in additional thermal emissions from the disk. During the fit, we left the disk density free to vary without imposing a link between the two components. The other parameters were set as in model A. Even in this case, we found a poor fit with $\chi^2/\rm{d.o.f.} = 694/367$.
%and hard power law photon index ($2.14$).\\ 
%We left the disk density free to vary, without imposing a link between the two components (best fit parameters: $19.0$ and $15.0$). 
%The other parameters were set as in first model, but in \emph{relxillD} and \emph{xillverD} the cut-off energy is fixed at $300\, \rm{keV}$.\\
For the residuals of the fit see Fig.~\ref{ress_riflessioneD}. With this model we found the following best fit parameters for the two densities: $\log N [\rm{cm^{-3}}] = 17.69^{+0.01}_{-0.03}$ and a lower limit at $\log N [\rm{cm^{-3}}]$ of 18.97.

In the relxillD framework, the maximum density considered is $10^{19} \rm{cm^{-3}}$. Therefore, we attempted to fit the data using a model incorporating a high-density component to characterize the reflection (\textsc{reflionxHD}), which supports densities up to $10^{22} \rm{cm^{-3}}$.
In this model (referred to as model C\footnote{model C:\textsc{const×tbabs×cloudy(\,cutoffpl\,+\,reflionxhc\,+\\+\,kdblur×reflionxHD)}}), we convolved reflionxHD with kdblur to account for relativistic effects. Instead of xillver, we utilized reflionxhc to describe the reflection, which calculates the reflected spectrum for optically-thick material of constant density. Additionally, we introduced a warm absorber (also using a \textsc{cloudy} table) in an attempt to better model the soft emission. However, we still obtained a poor fit ($\chi^2/\rm{d.o.f.} = 683/366$) with a density of $(7.64\pm1.22)\,\times\,10^{17}\rm{cm^{-3}}$.

%For this reasons we modelled the soft excess even phenomenologically, with a multicolour disk blackbody component (\emph{diskbb} in \emph{xspec}).\\
%The normalization and the inner disk temperature were left free to vary.\\
%In Fig.~\ref{mod_refl} and ~\ref{res_bad}  we shows - respectively - the model and the spectra  of the source; while all the best-fitting parameters are reported in Table~\ref{tab:4}.\\
%With this model we found $\chi^2/\rm{d.o.f.}=526/370$.\\

%\begin{figure}
%	\includegraphics[width=\columnwidth]{Figura4_new.png}
 %   \caption{Upper panel: Spectroscopic data and best fit model between 0.35 and 30 keV. Lower panel: residuals of the model. Model used during the fit: model B. \textit{XMM-Newton} data are in black, the \textit{NuSTAR FMPA} data in red and the \textit{NuSTAR FMPB} data in green.}
  %  \label{res_riflessioneD}
%\end{figure}

%\begin{figure}
%	\includegraphics[width=\columnwidth]{mod-refl.png}
%    \caption{Model used during the fit of \textit{XMM-Newton} and \textit{NuSTAR} data between 0.3 and 30 keV.}
%    \label{mod_refl}
%\end{figure}

%\begin{figure}
%	\includegraphics[width=\columnwidth]{res_bad.png}
 %   \caption{Upper panel: HE 1029-1401 spectra between 0.3 and 30 keV. Lower panel: residuals of the model. Model used during the fit: $\emph{const*phabs*(\,cutoffpl\,+\,relxill\,+\,xillver\,+\,diskbb)}$.}
 %   \label{res_bad}
%\end{figure}

%\begin{table}
%\centering
%\label{tab:4}
%\caption{\textit{XMM-Newton} and \textit{NuSTAR} data between 0.3 and 30 keV.\\
%Model used during the fit: $\emph{const*phabs*(\,cutoffpl\,+\,relxill\,+\,xillver\,+diskbb)}$.} 
%\begin{tabular}{ccc}
%\hline
%$kT_{\rm{bb}}$ & $\rm{eV}$ & $159_{-3}^{+4}$\\ %forse troppo alta%
%\rule[-5mm]{0mm}{1cm}
%$N_{bb}$ &  $\rm{ph}\,\rm{keV}^{-1}$ $\rm{cm}^{-2}$ $\rm{s}^{-1}$ & $538_{-60}^{+48}$\\
%\hline
%$N_{\rm{rel}} (10^{-5})$ &  $\rm{ph}\,\rm{keV}^{-1}$ $\rm{cm}^{-2}$ $\rm{s}^{-1}$ & $9.56_{-0.90}^{+0.27}$\\
%\hline
%$\rm{A}_{\rm{Fe}}$ & & $0.70\pm0.05$\\ %forse troppo bassa%
%\rule[-5mm]{0mm}{1cm}
%$N_{\rm{xill}} (10^{-14})$ &  $\rm{ph}\,\rm{keV}^{-1}$ $\rm{cm}^{-2}$ $\rm{s}^{-1}$ & $2.12$\\  
%c'è qualcosa che non va?%
%\hline
%$\Gamma$ & & $2.01\pm0.01$\\ 
%$N_{\rm{pow}} (10^{-3})$ &  $\rm{ph}\,\rm{keV}^{-1}$ $\rm{cm}^{-2}$ $\rm{s}^{-1}$ & $4.71\pm0.10$\\  
%\hline
%\end{tabular}
%\label{risultati}
%\end{table}

\subsection{Testing the two corona model}
In this section, we studied HE 1029-1401 using the two-corona model. In this model (model D), the soft X-ray spectrum is reproduced only by a thermally Comptonized continuum, modeled using \nthcomp\ in \emph{xspec}. 
The hard X-ray band is also modeled with \nthcomp, which parameterizes the high-energy roll-over using the electron temperature.\footnote{model D: \textsc{const×tbabs×(nthcomp+nthcomp+relxillCp+xillverCp)}}
In model D for the characterization of the reflection component, we replaced \relxill\ and \xillver\ with \relxillCp\ and \xillverCp, fitting for the electron temperature instead of the exponential cut-off.
We fixed the inclination angle at $30^{\circ}$, the outer radius of the accretion disk at $400\,\rm{R}_{\rm{g}}$, and the spin of the black hole at $0.998$. During the fit, we assumed that seed photons arise from a multicolor accretion disk, and all other parameters were left free to vary.\\
Using this approach, we obtained an acceptable fit with a $\chi^2 / \rm{d.o.f.}= 429/365$. In an attempt to represent the residuals in the soft band, we tried to include a warm absorber (modelled using the spectral synthesis code \textsc{cloudy}, as in the previous section) but no significant improvement was observed. 
We then extended our spectral investigation to the optical-UV
domain by including the OM data in the fit (model E).
\citeauthor{Petrucci1} (\citeyear{Petrucci1}) applied a model to the archival \textit{XMM-Newton}/OM and pn data of the source, incorporating two \textsc{nthcomp} components \citep{Zdz, Zycki} to represent the Comptonization continuum originating from the two coronae. Additionally, they included a single non-relativistic reflection component and a component to account for the contribution of the broad-line region (BLR) in the optical/UV band.\footnote{\textsc{tbabs × redden × mtable\{cloudytable\} × (nthcomp + nthcomp + atable\{smallBB\} + atable\{galaxy\} + xillver)}} In the same way of \citeauthor{Petrucci1} (\citeyear{Petrucci1}), we considered the potential influence of the broad-line region on the spectra including a model for the small blue bump, formed by the Balmer continuum toghether with Fe II emission. Our approach is based on an additive table with freely adjustable normalization (\textsc{smallBB})\footnote{The contribution by the host galaxy is not strong, being a factor of 10 below the nuclear flux both in the optical band \citep{Husemann} and in the UV band (estimated with GALFIT; K. K. Gupta, private communication). Given the coarse resolution of the OM photometric filters, and the fact that most of the filters used here are in the UV band, we opt to include only the BLR emission component, which dominates in the UV band over the galaxy \citep{2015Med}.}. 
A comprehensive description of this table for NGC 5548 is provided by \citet{2015Med}.
We, as well as \citeauthor{Petrucci1} (\citeyear{Petrucci1}), also included the effect of Galactic extinction (using the \textsc{redden} function in \emph{xspec}), with the reddening fixed during the fit at $E(B-V)=0.0954$ \citep{Guver}.\footnote{model E:\textsc{const×tbabs×redden×(nthcomp+nthcomp+smallBB+\\relxillCp+xillverCp)}}
The fit resulted in a chi-squared value of $\chi^2 / \rm{d.o.f.}= 437/368$. Specifically, the fit showed a not very much ionized accretion disk ($\log(\xi /\rm erg\,\rm cm\,\rm s^{-1})=1.30^{+0.24}_{-0.12}$), an internal radius of the accretion disk equal to $32_{-10}^{+23}\,\rm{R}_{\rm{g}}$, and a photon index and an electron temperature $\Gamma^{\rm{h}}=1.86 \pm 0.02$ and $kT_{\rm{e}}^{\rm{h}}$ = $17_{-2}^{+4}$ keV for the hot corona and $\Gamma^{\rm{w}}=2.75 \pm 0.05$ and $kT_{\rm{e}}^{\rm{w}}$ = $0.39_{-0.03}^{+0.04}$ keV for the warm corona.

\begin{figure}[htp]
\centering
\subfloat{\includegraphics[width = 0.4 \textwidth]{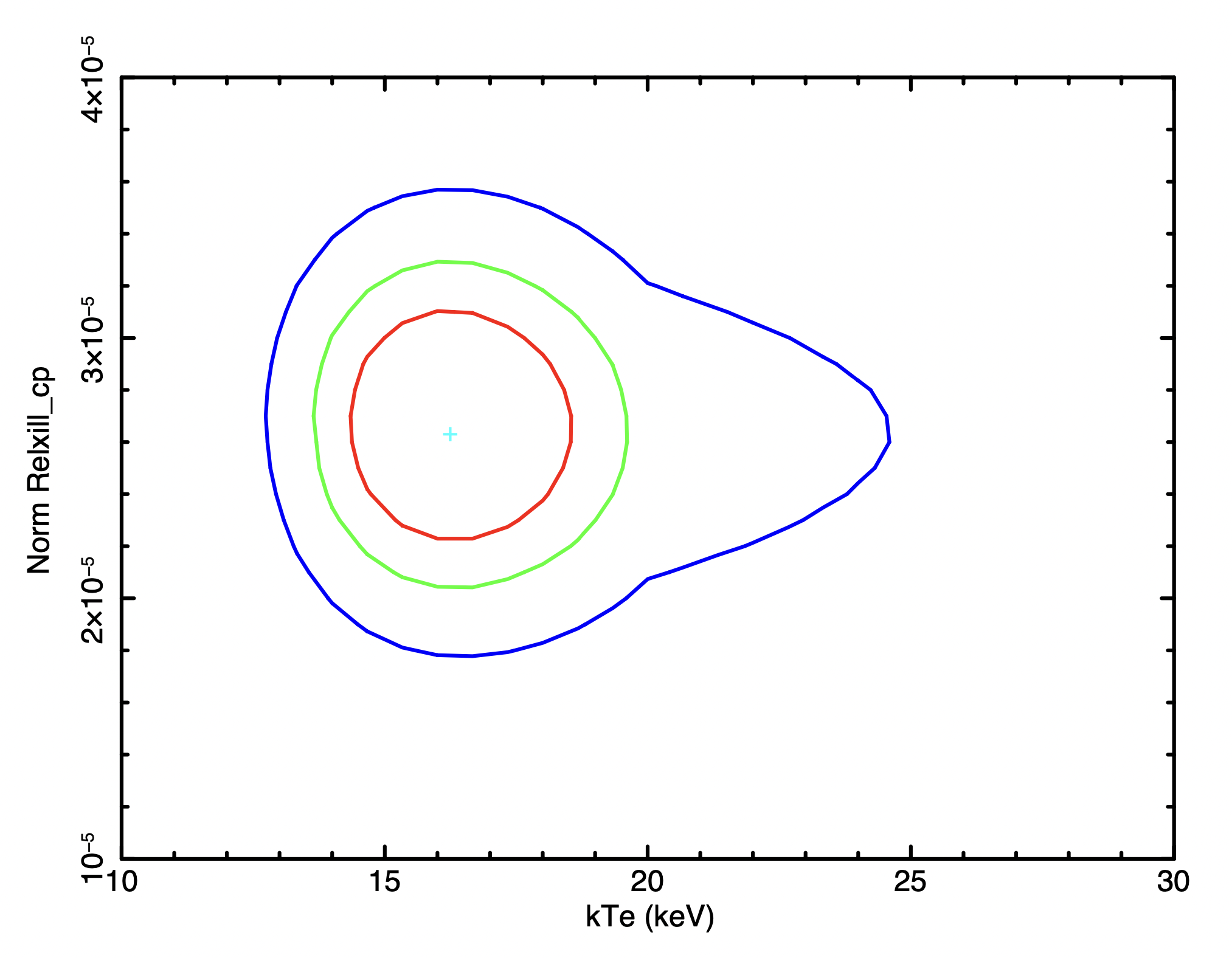}} \label{res_riflessioneD}
\subfloat{\includegraphics[width = 0.4 \textwidth]{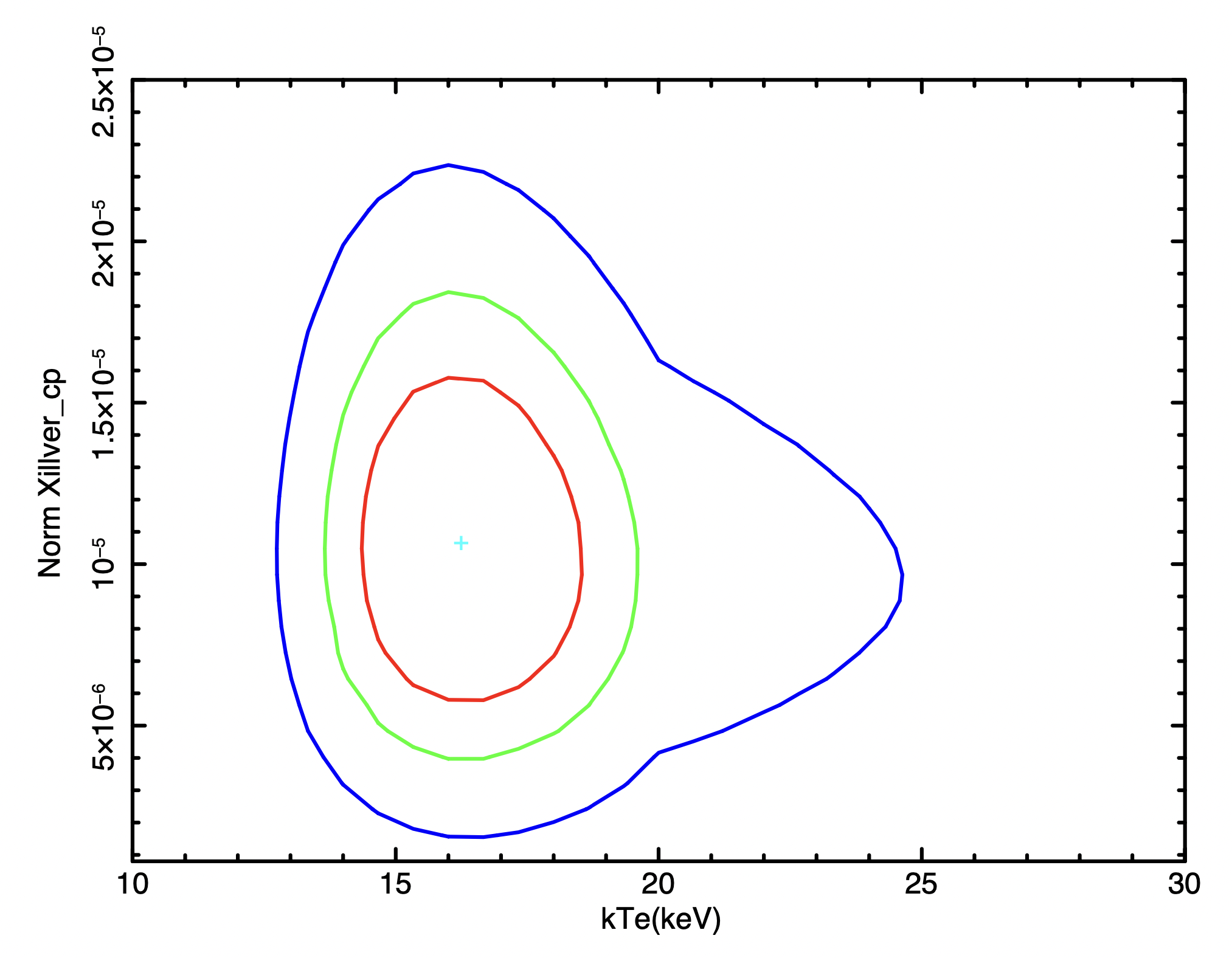}} \label{res_riflessioneD1}
\caption{Upper panel: Contour plot of \textsc{relxillCp} normalization versus electron temperature in the hot corona. Lower panel: Contour plot of \textsc{xillverCp} normalization versus electron temperature in the hot corona. The solid blue, green, and red curves represent the 68\%, 90\%, and 99\% confidence levels, respectively. Both plots are generated using model E.}
   \label{cp}
\end{figure}

\begin{figure}
	\includegraphics[width=\columnwidth]{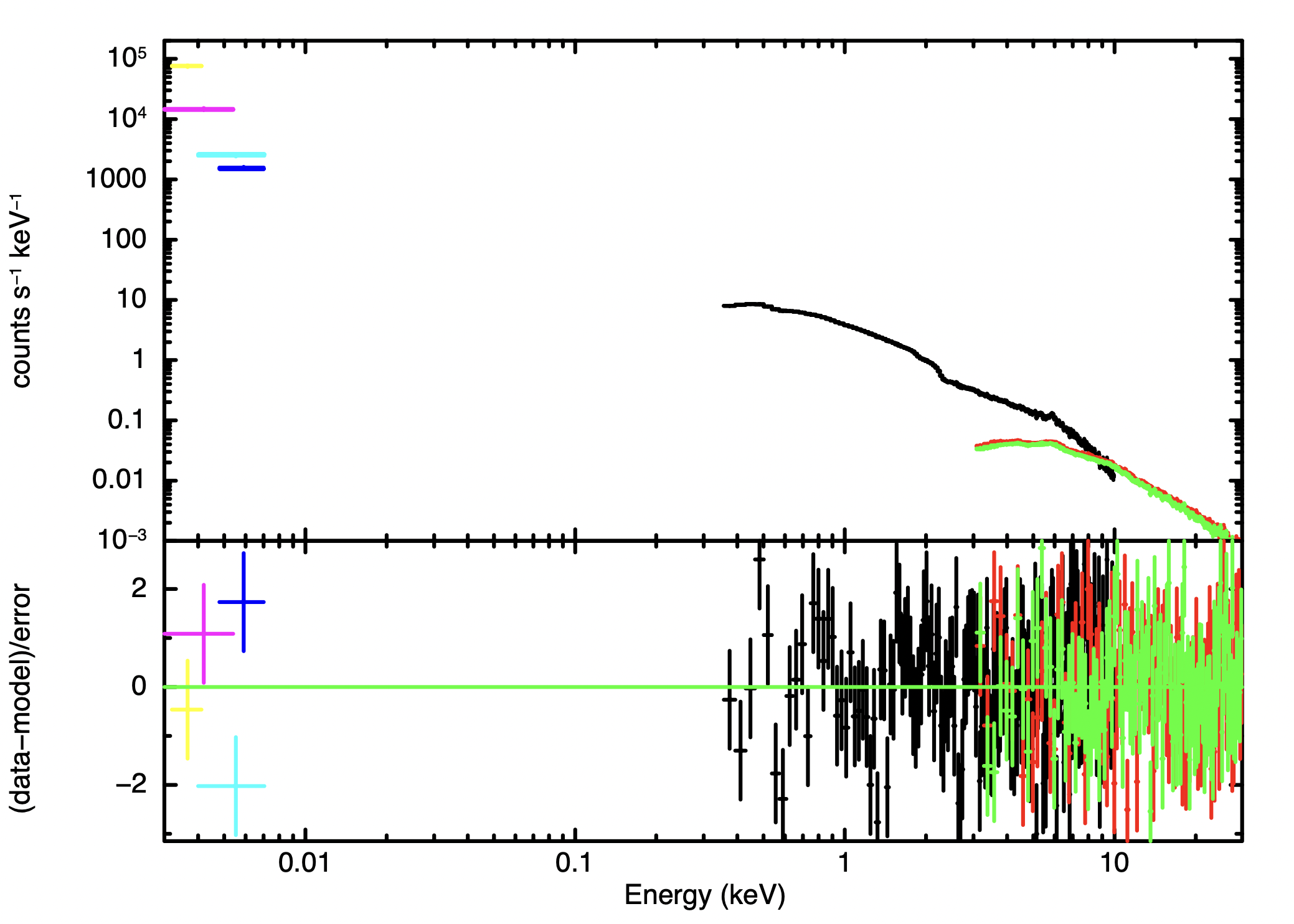}
    \caption{Upper panel: HE 1029-1401 data and best fit model. Lower panel: residuals of the model. Model used during the fit: model E. \textit{XMM-Newton} data are in black, the \textit{NuSTAR FMPA} data in red and the \textit{NuSTAR FMPB} data in green.}
    \label{res2}
\end{figure}

\begin{figure}
	\includegraphics[width=\columnwidth]{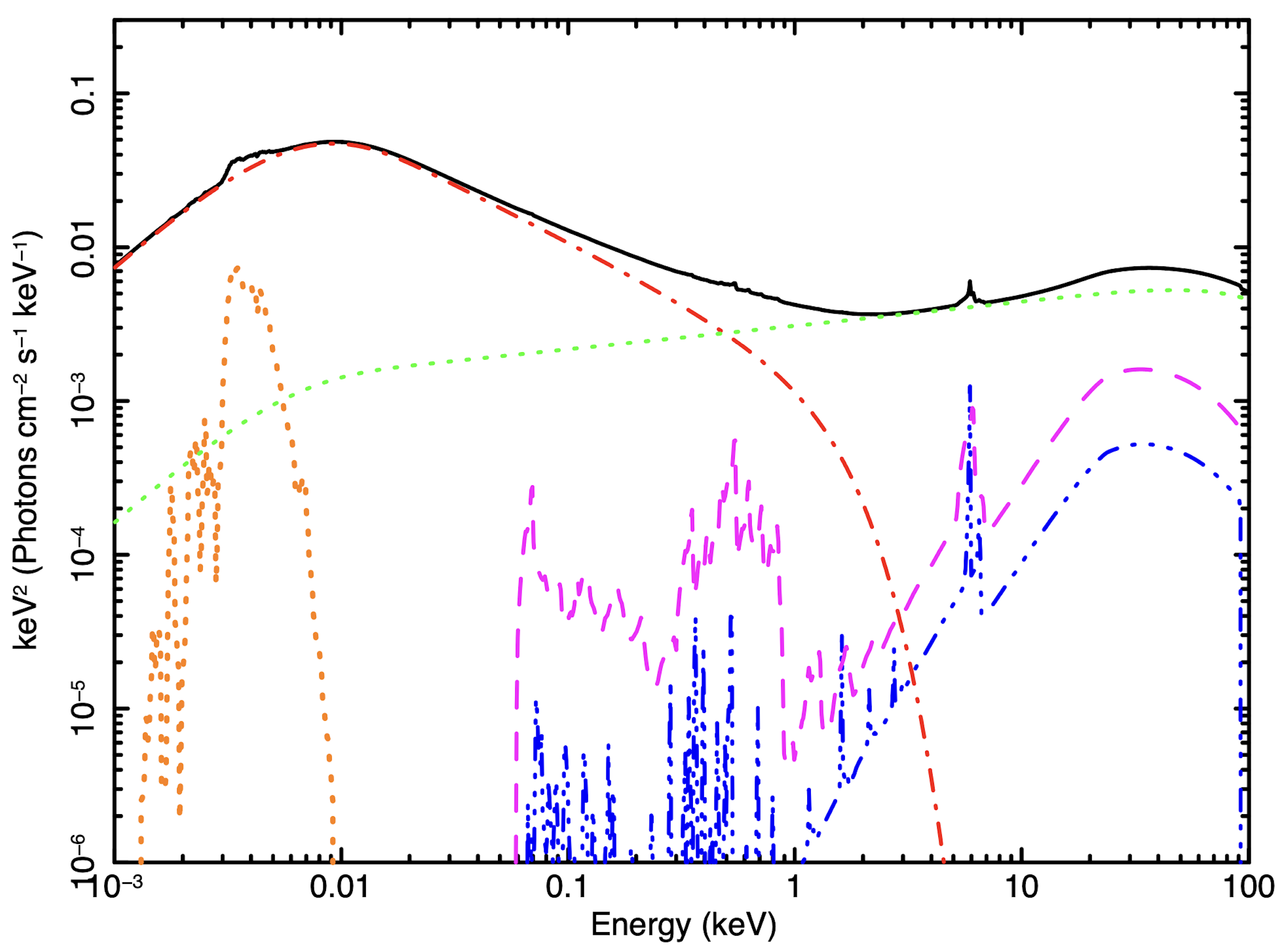}
    \caption{Best-fit model of the UV/X-ray spectrum of HE 1029-1401. The solid thicker black line is the best-fit model while the dashed lines are the different spectral components: in green the emission of the hot corona, in red the emission of the warm corona, the non-relativistic reflection component in blue, the relativistic reflection in pink and in orange the contribution of the small blue bump.}
    \label{model2}
\end{figure}

The optical depth of the corona is not one of the fit parameters but can be obtained, starting from the index $\Gamma$ and the temperature of the electrons, using the relation from \citet{1977Poz}:
\begin{equation}
    \tau_e=\sqrt{2.25+\frac{3}{\theta_e((\Gamma+0.5)^2-2.25)}}-1.5,
\end{equation}
where $\theta_e$ is the electron temperature normalized to the electron rest energy. For the hot and warm corona we find, respectively: $\tau^{\rm{h}}=\,5.4\pm0.9$, and  $\tau^{\rm{w}}=23\pm3$.\\
The fitting parameters are reported in Table~\ref{risultati}. Fig.~\ref{res2} shows the spectra and residuals of the model E, while in Fig.~\ref{model2} the best-fit model is represented.
According to this model, the broad-band spectrum of HE 1029-1401 is well described by two Comptonization processes. Specifically, the hard X-ray emission is primarily due to Comptonization in the hot corona, while the optical/UV to soft X-ray emission is dominated by Comptonization in a warm corona.
The potential degeneracy of the hot corona temperature with the reflection fraction has been investigated, yet no indications of degeneracy were observed (see Fig.~\ref{cp}).
The absorption-corrected model luminosity in the 0.001--1000 keV band is $6 \times 10^{45}$ erg~s$^{-1}$, corresponding to an Eddington ratio of $\sim$0.1.

\begin{table}
\centering
\caption{Best-fitting parameters of model E described in Sect. 3.3.}
\label{risultati}
\begin{tabular}{ccc}
\hline
\rule[-5mm]{0mm}{1cm}
$\log \xi_{i}$ & $\rm{erg}\,\rm{s}^{-1}$ $\rm{cm}$ & $1.30_{-0.12}^{+0.24}$\\
\rule[-5mm]{0mm}{1cm}
$\rm{A}_{\rm{Fe}}$ & & $2.2_{-0.6}^{+0.7}$\\
$N_{\rm{xill}} (10^{-5})$ & $\rm{ph}\,\rm{keV}^{-1}$ $\rm{cm}^{-2}$ $\rm{s}^{-1}$ & $1.0\pm0.5$\\
\rule[-5mm]{0mm}{1cm}
$R_{\rm{int}}$ & $\rm{R}_{\rm{g}}$ & $32_{-10}^{+23}$\\
\rule[-5mm]{0mm}{1cm}
$N_{\rm{rel}} (10^{-5})$ & $\rm{ph}\,\rm{keV}^{-1}$ $\rm{cm}^{-2}$ $\rm{s}^{-1}$ & $2.7\pm0.6$\\
\hline
\rule[-5mm]{0mm}{1cm}
$N_{\rm{small-bb}} (10^{-2})$ & $\rm{ph}\,\rm{keV}^{-1}$ $\rm{cm}^{-2}$ $\rm{s}^{-1}$ & $1.8\pm0.9$\\
\hline
\rule[-5mm]{0mm}{1cm}
$\Gamma^{\rm{w}}$ & & $2.75 \pm 0.05$\\ 
\rule[-5mm]{0mm}{1cm}
$kT_{\rm{e}}^{\rm{w}}$ & $\rm{keV}$ & $0.39_{-0.03}^{+0.04}$\\
\rule[-5mm]{0mm}{1cm}
$kT_{\rm{bb}}^{\rm{w}}$ & $\rm{eV}$ & $2.9 \pm 0.5$\\
\rule[-5mm]{0mm}{1cm}
$N_{\rm{NthComp}}^{\rm{w}} (10^{-3})$ & $\rm{ph}\,\rm{keV}^{-1}$ $\rm{cm}^{-2}$ $\rm{s}^{-1}$ & $1.39_{-0.11}^{+0.13}$\\
\hline
\rule[-5mm]{0mm}{1cm}
$\Gamma^{\rm{h}}$ & & $1.86 \pm 0.02$\\
\rule[-5mm]{0mm}{1cm}
$kT_{\rm{e}}^{\rm{h}}$ & $\rm{keV}$ & $17_{-2}^{+4}$\\
\rule[-5mm]{0mm}{1cm}
$N_{\rm{NthComp}}^{\rm{h}} (10^{-3})$ & $\rm{ph}\,\rm{keV}^{-1}$ $\rm{cm}^{-2}$ $\rm{s}^{-1}$ & $3.81_{-0.12}^{+0.11}$\\
\hline
\rule[-5mm]{0mm}{1cm}
$F_{0.5-2} (10^{-12})$ & $\rm{erg}s$ $\rm{s}^{-1}$ $\rm{cm}^{-2}$ & $8.04_{-0.02}^{+0.03}$\\       
\rule[-5mm]{0mm}{1cm}
$F_{2-10} (10^{-11})$ & $\rm{erg}s$ $\rm{s}^{-1}$ $\rm{cm}^{-2}$ & $1.07\pm0.01$\\
\rule[-5mm]{0mm}{1cm}
$L_{0.5-2} (10^{43})$ & $\rm{erg}s$ $\rm{s}^{-1}$ & $1.49\pm0.01$\\    
\rule[-5mm]{0mm}{1cm}
$L_{2-10} (10^{44})$ & $\rm{erg}s$ $\rm{s}^{-1}$ & $1.94\pm0.01$\\
\hline
\end{tabular}
\end{table}

As a further step in model F, we replaced the warm \nthcomp\ component with a table model computed by \cite{2020Petrucci}, who performed numerical simulations of the spectra of warm coronae using the radiative transfer code TITAN coupled with the Monte-Carlo code NOAR \citep{Dumont}.
This model, in its current implementation, assumes a black body seed photon spectrum, instead of a disk black body assumed in \nthcomp.\footnote{model F: \textsc{const×tbabs×redden×(nthcomp+titan+smallBB+\\relxillCp+xillverCp)}}
The free parameters of the TITAN table are the black body temperature of the seed photons, the heating power $q_h$ in the warm corona, the optical depth $\tau^{\rm{w}}$, and the normalization. 
During this fit, we left all the parameters of the TITAN table free to vary.
%, except obviously the redshift of the source. 
The temperature of the seed photons in the \nthcomp\ model that describes the hot corona was linked to that of the TITAN table.
%with appropriate changes of coordinates.\\ 
With this model, we found a chi-squared value of $\chi^2 / \rm{d.o.f.}= 443/368$ and the following best fit parameters for the warm corona. For the logarithm of the heating power $\log q_h$ (erg $\rm{s}^{-1}$ $\rm{cm}^3$) we found $-22.862_{-0.004}^{+0.002}$, for the logarithm of $T_{bb}$ (in unit of K) $4.38\pm0.05$, for the optical depth $\tau^{\rm{w}} = 18.4_{-0.6}^{+0.4}$ and the normalization was $1.53_{-0.16}^{+0.13}\times 10^{-4}$. Given the source distance and luminosity, from the model we can also estimate the radial size of the warm corona. The normalization of the table is indeed directly related to this quantity, with some simple counts (see \citealt{2020Petrucci}) we obtained $\sim 20\,\rm{R_g}$.\\
%We note that the TITAN model, in its current implementation, assumes a black body seed photon spectrum, which is not the same as the disk black body assumed in \nthcomp. However, the model provides a good fit to the data.\\

Finally in model G, we tested the model \textsc{reXcor} by \cite{xiang2022} \citep[see also][]{ballantyne2020,2024Ballantyne}, which incorporates both warm Comptonization and ionized reflection from a warm corona illuminated by a lamp-post hot corona. In this model, the accretion energy is distributed between the disk, the warm corona and the hot corona. The relative strength of warm Comptonization and of ionized reflection is driven by the heating fraction of the warm corona and of the hot corona, respectively. The model does not include optical/UV emission, thus for simplicity we fitted only the X-ray spectra. Since \textsc{reXcor} includes blurred ionized reflection, we replaced both the warm Comptonization component and the \textsc{relxill} reflection component with \textsc{reXcor}.\footnote{model G: \textsc{TBfeo×const×(xillverCp+reXcor+nthcomp)}} We kept the \textsc{xillvercp} component to reproduce distant reflection. In \textsc{reXcor} we assumed a lamp-post height of 20 $\rm{R_g}$, a black hole spin of 0.99, and an Eddington ratio of 0.1 consistent with the results reported by \cite{2024Ballantyne}. We initially found a poor fit in the soft band with strong residuals near 0.5 keV. To account for these, we included two narrow Gaussian absorption lines, which are found to be at 0.62 keV and 0.74 keV. Gaussian lines of this type, here briefly described, are often needed in this kind of fit, see for example \cite{2024Ballantyne}.  We obtained an acceptable fit with $\chi^2 / \rm{d.o.f.}= 436/364$ and the following best-fit parameters for \textsc{reXcor}: hot corona heating fraction $f_X = 0.030^{+0.015}_{-0.003}$, warm corona heating fraction $0.40^{+0.05}_{-0.03}$ and optical depth $\tau < 11$ (the lower limit of the model is 10).

\section{Discussion and conclusions}

The origin of the soft excess in the  X-ray spectra of AGN is still debated. Here we have presented the analysis of XMM-Newton and NuSTAR simultaneous observation of the nearby (z=0.086) AGN HE 1029-1401 with the aim of revealing the nature of the soft-excess in this source. Our findings can be summarized as following: 

\begin{itemize}
\item The spectrum of the source shows the presence of a narrow Fe K$\alpha$ line and a broad ($\sigma=0.38_{-0.16}^{+0.26}$ keV)  Fe K$\alpha$ line, at $6.4$ keV.
\item To account for emissions resulting from reflection, it was essential to incorporate both a relativistic and a non-relativistic component into the best fit model.
\item The study confirms the presence of a significant soft X-ray excess below $2$ keV. No warm absorber has been detected.
\item We find that the two corona model describes the source better than the reflection model. With a phenomenological model we measure for the hot corona a relatively low temperature $kT_{\rm{e}}^{\rm{h}}$ = $17_{-2}^{+4}$ keV, and an optical depth of $\tau^{\rm{h}}=\,5.40\pm0.85$ (assuming a spherical geometry); for the warm corona, we obtain parameters consistent with lower luminosity Seyferts: $\Gamma^{\rm{w}}=2.75\pm0.05$, $kT_{\rm{e}}^{\rm{w}}$ = $0.39_{-0.04}^{+0.06}$ keV, $\tau^{\rm{w}}=23\pm3$.
\item We have also tested two more physical models for the warm corona, namely the TITAN-NOAR table model by \cite{2020Petrucci}, and the \textsc{reXcor} model by \cite{xiang2022}. 
Despite the different physical assumptions \cite[for a more detailed comparison, see][]{2020Petrucci}, both models provide a good fit to the data. The results are consistent with the presence of a warm corona with an optical depth of the order of 10--20 and with a radial extension of a few tens of gravitational radii.
%We have achieved a good fit comparable to that obtained using Nthcomp. Our findings suggest that the TITAN tables may be a useful tool for modeling the warm coronae spectra. 
\end{itemize}

%\blue{problem: TITAN and reXcor give different optical depth, by almost a factor of 2 (18 vs 10). Also the radius Rwarm is 20 Rg from TITAN and 40-80 Rg from reXcor (Fig. 2 of Xiang+22)}

This is the first \textit{NuSTAR} observation of the source. However, previous X-ray  observations of this source were conducted using the \textit{XMM-Newton} telescope and were analyzed in a previous paper by \citet{Petrucci1}. The results for the warm corona reported in the current study are comparable with those reported in the previous paper indicating consistency between the two studies. Specifically, for the hot corona we found slightly different parameters due to a different assumption in the temperature of the hot electrons.
 The results obtained in Model G can be compared with the analysis of another observation of the source using the same model as reported in \cite{2024Ballantyne}. Our work yields a hot corona heating fraction that is compatible with that of \cite{2024Ballantyne}, alongside a lower value for the warm corona heating fraction and a slightly larger photon index. We also computed the ratio of the 0.3–10 keV \textsc{reXcor} and primary continuum fluxes; we obtain a value of 0.36, compatible with \cite{2024Ballantyne}. Our results are overall consistent with the trends between the \textsc{reXcor} parameters and the Eddington ratio discussed by  \cite{2024Ballantyne}, see their Fig. 1--4.\\ We can compare the results obtained with an other quasar for which the two coronae model was tested: RBS 1055 \citep{Marinucci2}. In our study, we have observed that the coronae of RBS 1055 are hotter compared to those found for HE 1029-1401 in this study. This could be due to a number of factors such as differences in the physical properties of the black hole or the surrounding environment, or variations in the accretion rate and efficiency. Indeed the Eddington ratio of HE 1029-1401 is a factor two larger than the Eddington ratio of RBS 1055 ($L_{\rm{bol}}/L_{\rm{Edd}} \simeq 0.05$).  

In a wider scenario, the values obtained from this study align with those known for radio-quiet AGNs described by the two coronae model (see \citealt{Petrucci1}): a warm corona, with an optical depth $\tau \simeq 10 - 40$ and an electron temperature $\rm{kT} \simeq 0.1 - 1$ keV and a hot corona, optically thin ($\tau \simeq 1$) and hot ($\rm{kT} > 20$ keV).\\

In conclusion, the two-corona model provides a good fit of the data, indicating that warm Comptonization is a likely explanation for the soft excess. The broad-band UV/X-ray properties of this luminous quasar are overall consistent with those of local Seyferts, hinting at a similar structure of the accretion flow. This is broadly consistent with the results of \citealt{Mitchell}, who analyse the spectral energy distribution of a large sample of quasars at $z\leq 2.5$ and find that the spectral shape of the optical-UV continuum remains nearly constant across decades of black hole mass for a given luminosity. \citealt{Mitchell} point out that standard accretion disc models cannot reproduce this behavior, which can instead be explained if the disc is completely covered by a warm Comptonising corona, as long as the UV spectral shape correlates with the accretion rate (which is indeed observed, see \citealt{Petrucci1}). In our case, we find a relatively steep photon index of the warm corona, in agreement with the Eddington ratio measured for HE1029-1401.

The existence of the warm corona could yield significant implications, such as directly influencing our comprehension of the vertical equilibrium of the accretion disk. Moreover, models E, F, and G provide a good description of the X-ray  spectrum, but their distinct assumptions highlight the necessity for further investigations to enhance our comprehension of warm coronal properties. Future observations of luminous quasars will enable further studies of the soft excess in different accretion regimes.

\section*{Data Availability}

The data analysed in this paper are publicly available via the \xmm\ and \nustar\ archives. See Table 1 for the Obs. ID and the details of the observations.

\begin{acknowledgements}

We thank the referee for useful comments that improved the manuscript.
This research was supported by the International Space Science Institute (ISSI) in Bern, through ISSI International Team project \#514 (Warm Coronae in AGN: Observational Evidence and Physical Understanding). 

The research leading to these results has received funding from the
European Union’s Horizon 2020 Programme under the AHEAD2020 project (grant
agreement n. 871158).

This publication was produced while B.V. attending the PhD program in  in Space Science and Technology at the University of Trento, Cycle XXXVIII, with the support of a scholarship financed by the Ministerial Decree no. 351 of 9th April 2022, based on the NRRP - funded by the European Union - NextGenerationEU - Mission 4 "Education and Research", Component 1 "Enhancement of the offer of educational services: from nurseries to universities” - Investment 4.1 “Extension of the number of research doctorates and innovative doctorates for public administration and cultural heritage”.
POP acknowledges financial support from the french space agency (CNES) and the High Energy National Programme (PNHE) from CNRS.

\end{acknowledgements}

\bibliography{example}
\bibliographystyle{aa}

\end{document}